\documentclass[a4paper,12pt]{article}
\usepackage{amsmath}
\usepackage{amsfonts}
\usepackage{amssymb}
\usepackage{graphicx}
\usepackage{cite}

\begin{document}

\date{}

\title{\bf Anisotropic solutions in $f(R)$ Gravity}

\author{S. K. Tripathy\footnote{Department of Physics, Indira Gandhi Institute of Technology, Sarang, Dhenkanal, Odisha-759146, INDIA, tripathy\_ sunil@rediffmail.com}
and
B. Mishra 
\footnote{Department of Mathematics, Birla Institute of Technology and Science-Pilani,Hyderabad Campus,Hyderabad-500078, INDIA, bivudutta@yahoo.com}    
}

\maketitle

\begin{abstract} Anisotropic cosmological models are investigated in the frame work of $f(R)$ gravity in the metric formalism. Plane symmetric models are considered to incorporate anisotropy in the expansion rates along different spatial directions.  The anisotropy in expansion rates are assumed to be maintained throughout the cosmic evolution. Two accelerating models are constructed by considering different functional forms for $f(R)$. The viability of these models are tested through a stability analysis.
\end{abstract}



\section{Introduction}

Cosmological research in recent times mostly centres around the fact that the universe is in a state of accelerated expansion. This late time cosmic dynamics has been confirmed from a lot of observations  \cite{Riess98, Perl99, Spergel07, Komatsu09, Seljak05, Eisen05}. It is believed that the transition from a decelerated phase to an accelerated one may have occurred at a transition redshift $z_{da}\sim 1$ \cite{Farooq13, Copo14}. The late time cosmic acceleration has trigged a lot of research interest but its cause is not yet known exactly. Two different approaches are adopted to address this issue. As a first approach, the cosmic speed up is explained in the framework of General Relativity through the inclusion of an exotic dark energy in the matter field. The dark energy, usually represented by a cosmological constant, corresponds to an isotropic fluid with almost constant energy density with negative pressure.  Many alternative dark energy models have also been proposed in recent times: canonical scalar fields like quintessence \cite{Ratra88,Sahni00}, phantom fields \cite{Cald02}, k-essence \cite{Picon00, Picon01}, tachyons \cite{Sen02}, quintom \cite{Feng05, Guo05}; parametrised dark energy candidates such as ghost dark energy \cite{Urban09, Urban10, Ohta11}, holographic dark energy \cite{Li04},  Ricci dark energy \cite{Cao09} and agegraphic dark energy \cite{Cai07, Wei08} or a consideration of a unified dark fluid \cite{Anand06, Xu12, SKT15a}. Dark energy is required to explain the cosmic speed up and dark matter is required to explain the emergence of Large Scale Structure in the universe. According to the Planck data \cite{Ade14, Ade14a, Ade14b, Ade15}, the dark energy leads the cosmic mass-energy budget with a share of $\Omega_{\Lambda}=0.691\pm 0.006$ followed by dark matter having $\Omega_{dm}=0.259\pm 0.005$. However, the lack of a comprehensive understanding of the nature and behaviour of these components either in particle form or of scalar fields has triggered to look for alternative theories of gravity \cite{Martino15}. Further, the need of two unknown components to explain physical phenomena can be interpreted as a break down of the very theory at Astrophysical scales \cite{Martino15, Capo11}. 

Another approach is the  modification of the geometrical part of the gravitational interaction by including higher order curvature invariants in the Einstein-Hilbert action. Among various ways to modify the Einstein's General Relativity, a much straightforward approach is the $f(R)$ theory, where the Ricci curvature scalar $R$ in the action is replaced by a more general function of $R$. After the first proposal by Buchdal \cite{Buchdal70}, the theory has been developed further by others \cite{Breiz71, Star80, Capo03, Nojiri03, Caroll04, Nojiri06}. Of late, in the context of late of time cosmic speed up, the $f(R)$ theory of gravity has generated a great deal of research interest \cite{Capo11, Nojiri11, Olmo11, Satiriou07, Satiriou10, Felice10 }. $f(R)$ gravity models have been widely used with emphatic results and success in stellar formation and evolution \cite{Asta15, Asta15a, Lau13}, structure formation and evolution of the universe \cite{Abdel08, Carloni09, Nojiri14, Hu15, Bamba15}, cluster of galaxies \cite{Schmidt09, Ferraro11, Teru12}, gravitational waves and massive gravitons \cite{Anton13, Clifton10, Lau11}.

The universe is mostly observed to be flat and isotropic and can be well described by $\Lambda$ dominated Cold Dark Matter($\Lambda$CDM) model.  The Cosmic Microwave Background (CMB) angular power spectrum of perturbation is in excellent agreement with the predictions of the standard model. However, despite the success of $\Lambda$CDM model, observations of high resolution CMB radiation data from Wilkinson Microwave Anisotropy Probe (WMAP) provide some of its anomalous features at large scale. Precise measurements of WMAP showed that the quadrupole $C_2$ and Octupole $C_3$ are usually aligned and are concentrated in a plane about $30^0$ to the galactic plane \cite{Tegmark03, Costa04}. These observations  suggest a non-trivial topology of the large scale geometry of the universe with an asymmetric expansion \cite{Hinshaw09, Costa04, Watan09}. Planck data also show a slight redshift of the primordial power spectrum of curvature perturbation from exact scale invariance \cite{Ade14} that suggests the presence of some anisotropic energy source in the universe. While there are many ways suggested to handle the issue of global anisotropy, the flat Friedman model (FRW) can be suitably modified to incorporate the anisotropic effect. In recent times, some anisotropic models bearing similarity to Bianchi morphology have been proposed \cite{Campa06, Campanel07, Gruppo07, Jaffe05, Jaffe06a}. In this context, it is interesting to investigate some anisotropic models in the framework of $f(R)$ gravity. The importance of the effects of anisotropy in the universe has led many authors to investigate anisotropic models with or without matter in the field equations \cite{Shamir10, Sharif11, Aktas12, Yilmaz12, Singh13}.

Motivated by the earlier works, we have investigated some anisotropic models in the framework of $f(R)$ gravity. We have obtained vacuum solutions to the field equations. From certain plausible physical basis, we have constructed some $f(R)$ models which are stable and can be further used to study some other effects. The paper is organised as follows: In Section 2, the basic formalism of the $f(R)$ gravity is presented in brief. We have not considered any matter field and restrict ourselves to vacuum solutions only.  We have considered an anisotropic universe described through a Locally Rotationally Symmetric Bianchi type -I metric (LRSBI) to handle the anisotropic effects. LRSBI models are the generalisation of FRW models with asymmetric expansions along different spatial directions. The dynamics of LRSBI universe in the $f(R)$ gravity are given in a most general form in Section-3. The field equations are simplified by assuming that the shear scalar and the expansion scalar are proportional to each other. This assumption leads to an anisotropic relationship among the directional scale factors. The incorporated anisotropy in the expansion rates are considered to be maintained through out the cosmic evolution. In Section-4, two classes of cosmological models have been constructed by considering different functional forms of $f(R)$. The model parameters are constrained from certain physical basis. At the end, the conclusions of the work are presented in Section 5.

\section{Basic Formalism}

The four dimensional modified Einstein-Hilbert action in $f(R)$ gravity is taken as
\begin{equation}
S=\frac{1}{2\kappa^2}\int d^4x \sqrt{-g}f(R)+ \int d^4x \sqrt{-g}\mathcal{L}_{m},\label{eq:1}
\end{equation}
where $f(R)$ is an arbitrary function of the Ricci scalar $R$;  $R=g^{\mu \nu}R_{\mu \nu}$. $R_{\mu \nu}$ is the Ricci tensor, $\mathcal{L}_{m}$ is the matter Lagrangian, $g$ is the determinant of the metric $g_{\mu \nu}$ and $\kappa^2=\frac{8\pi G}{c^4}$. $G$ is the Newtonian Gravitational constant and $c$ is the speed of light in vacuum. Here we choose the unit system where $8\pi G=c=1$.  The gravitational field equations for $f(R)$ gravity can be obtained by varying the action with respect to the metric tensor $g_{\mu \nu}$ as 
\begin{equation}
f_RR_{\mu \nu}-\frac{1}{2}f(R)g_{\mu \nu}-\bigtriangledown_{\mu}\bigtriangledown_{\nu}f_R+g_{\mu \nu}\Box f_R= \kappa T_{\mu \nu},\label{eq:2}
\end{equation}
where $f_R=\frac{df(R)}{dR}$ and $\Box\equiv\bigtriangledown^{\mu}\bigtriangledown_{\mu}$ is the de Alembert's operator. $\bigtriangledown_{\mu}$ is the covariant derivative and $T_{\mu \nu}=-\frac{2}{\sqrt{-g}}\frac{\delta \left(\sqrt{-g}\mathcal{L}_{m}\right)}{\delta g^{\mu \nu}}$ is the energy momentum tensor corresponding to the matter Lagrangian $\mathcal{L}_{m}$. The above equation reduces to the usual field equations in General Relativity for $f(R)=R$.
Contraction of eq.\eqref{eq:2} yields
\begin{equation}
f_RR-2f(R)+3\Box f_R=T.\label{eq:3}
\end{equation}
Here, $T=g^{\mu \nu}T_{\mu \nu}$ is the trace of the energy momentum tensor. For a vacuum solution with $T=0$, the above equation reduces to a nice relationship between $f(R)$ and $f_R$ as
\begin{equation}\label{eq:4}
f(R)=\frac{1}{2}\left[3\Box +R\right]f_R.
\end{equation}
Also, for vacuum solutions, eq.\eqref{eq:2} can be rewritten as
\begin{equation}
\frac{1}{g_{\mu \nu}}\left[f_RR_{\mu \nu}-\bigtriangledown_{\mu}\bigtriangledown_{\nu}f_R\right]=\frac{1}{2}f(R)-\Box f_R.\label{eq:5}
\end{equation}
Using eq.\eqref{eq:4}, eq.\eqref{eq:5} is reduced to 
\begin{equation}
\frac{1}{g_{\mu \nu}}\left[f_RR_{\mu \nu}-\bigtriangledown_{\mu}\bigtriangledown_{\nu}f_R\right]=\frac{1}{4}\left[R-\Box\right]f_R.\label{eq:6}
\end{equation}
It is interesting to note here that, the right hand side of the above equation \eqref{eq:6} does not involve any indices like $\mu$ and $\nu$. Therefore, the diagonal elements for the left hand side are the same being independent of indices.

\section{Dynamics of an anisotropic universe in $f(R)$ gravity}
We consider an anisotropic LRSBI universe in the form
\begin{equation}
ds^2=dt^2-A^2(t)dx^2-B^2(t)\left(dy^2+dz^2\right).\label{eq:7}
\end{equation}
The model is plane symmetric with asymmetric expansion along the symmetry axis and symmetry plane. The asymmetry in expansion is considered through a simple assumption of different scale factors along different directions. The scale factor along the symmetry axis $A$ is chosen to be different from the scale factor $B$ along the symmetry plane. The directional scale factors $A$ and $B$ are considered to be functions of cosmic time only.

The Ricci scalar $R$ for LRSBI metric is given by
\begin{equation}
R=-2\left[\frac{\ddot{A}}{A}+2\frac{\ddot{B}}{B}+2\frac{\dot{A}}{A}\frac{\dot{B}}{B}+\left(\frac{\dot{B}}{B}\right)^2\right],
\end{equation}
where a dot over a variable denotes differentiation with respect to cosmic time.
The directional Hubble rates for LRSBI model are defined as $H_x=\frac{\dot{A}}{A}$ and $H_y=H_z=\frac{\dot{B}}{B}$, so that the mean Hubble rate can be $H=\frac{1}{3}\left(H_x+2H_y\right)$. The scalar expansion $\Theta$ and the shear scalar $\sigma$ are given as
\begin{eqnarray}
\Theta=3H=H_x+2H_y,\\
\sigma^2=\frac{1}{3}\left(H_x-H_y\right)^2
\end{eqnarray}
The average anisotropic parameter is given by 
\begin{equation}
\mathcal{A}_m=\frac{1}{3}\sum_{i=x,y,z} \left(\frac{H_i-H}{H}\right)^2
\end{equation}
Ricci scalar can be written in terms of the directional Hubble rates as 
\begin{equation}
R=-2\left[\dot{H_x}+2\dot{H_y}+H_x^2+3H_y^2+2H_xH_y\right].
\end{equation}
The field equations for LRSBI model, along with eq. \eqref{eq:6} can now be written as
\begin{eqnarray}
-2\left(\dot{H_y}+H_y^2\right)+H_x\left(2H_y+\frac{\dot{f_R}}{f_R}\right)-\frac{\ddot{f_R}}{f_R}=0,\label{eq:13}\\
-\left(\dot{H_x}+\dot{H_y}\right)-H_x^2+H_y\left(H_x+\frac{\dot{f_R}}{f_R}\right)-\frac{\ddot{f_R}}{f_R}=0.\label{eq:14}
\end{eqnarray}
In order to solve the above equations we require some additional conditions. The scalar expansion and shear scalar can be taken to be proportional to each other which provides us an anisotropic relation between the directional Hubble rates as $H_x=kH_y$ \cite{SKT10}. Here, $k$ takes care of the anisotropic nature of the model in the sense that, if $k=0$, we get an isotropic universe, otherwise, the model is anisotropic. It is required that, the anisotropic parameter $k$ should be a positive constant throughout the cosmic evolution. The Ricci scalar becomes
\begin{equation}
R=-2\left[(k+2)\xi \dot{H}+(k^2+2k+3)\xi^2 H^2\right],\label{eq:15}
\end{equation}
where, $\xi=\frac{3}{k+2}$ and is a positive constant taking care of the anisotropic nature of the model. For isotropic model, $\xi=1$.

The field equations \eqref{eq:13} and \eqref{eq:14} can now be expressed as
\begin{eqnarray}
-2(\dot{H}+\xi H^2)+kH\left[2\xi H+\frac{\dot{f_R}}{f_R}\right]-\frac{1}{\xi}\frac{\ddot{f_R}}{f_R}=0,\label{eq:16}\\
(k+1)\dot{H}+\xi k^2H^2-H\left[\xi kH+\frac{\dot{f_R}}{f_R}\right]+\frac{1}{\xi}\frac{\ddot{f_R}}{f_R}=0.\label{eq:17}
\end{eqnarray}

The above equations reduce to      
\begin{equation}
2\dot{H}-H\frac{\dot{f_R}}{f_R}+\frac{\ddot{f_R}}{f_R}=0,\label{eq:17a}
\end{equation}
for a flat isotropic model with equal rate of expansion in all spatial directions. One should note that for an isotropic model, the anisotropic parameters $k$ and $\xi$ reduce to 1. But $k$ is else than 1 for anisotropic models and  eq.\eqref{eq:17a} splits into the equations \eqref{eq:16} and \eqref{eq:17}. For anisotropic models, we can add eqs.\eqref{eq:16} and \eqref{eq:17} to get

\begin{equation}
\dot{H}+3H^2+H\frac{\dot{f_R}}{f_R}=0.\label{eq:18}
\end{equation}
Equation \eqref{eq:18} is an important relation that contains the essence of the functional $f(R)$ in LRSBI model. In General Relativity, $f(R)=R$ and therefore $\dot{f_R}$ vanishes. In this context, eq. \eqref{eq:18} reduces to $\dot{H}+3H^2=0$ which provides a decelerating universe. In other words, if the anisotropic relationship among the directional Hubble rates are presumed to be maintained through out the cosmic evolution in LRSBI model, we can not get accelerating models in General Relativity. This fact has already been investigated earlier \cite{SKT15a, SKT13, SKT14, SKT15b}. However, if we modify the Einstein-Hilbert action with a suitable functional form for $f(R)$ other than the linear one, $f(R)=R$, we may get viable accelerating models. In the later case, the contribution coming from the third term of eq. \eqref{eq:18} i.e. $\frac{\dot{f_R}}{f_R}$ necessarily helps in the cosmic acceleration. One can conceive this geometrical feature as the inherent dark energy contribution or that of a scalar field.
\section{Anisotropic models}
In principle, one can integrate eq. \eqref{eq:18} and by the use of some plausible assumptions on the functional form of $f(R)$, different cosmological models may be constructed for suitable applications in cosmology and Astrophysics. The constructed models may be tested for their consistency and viability in the context of observations. In the present work, we have constructed two classes of cosmological models by considering two different functional forms for $f(R)$. In constructing the models, we have adopted a view that, the first derivative $f_R$ of the functional $f(R)$ depends on the mean Hubble rate.

\subsection{Model-I}
Let us now assume that,
\begin{equation}
f_R= H^{\alpha},\label{eq:19}
\end{equation}
so that, $\frac{\dot{f_R}}{f_R}= \alpha \frac{\dot{H}}{H}$. Here $\alpha$ is an arbitrary constant parameter of the model.
With eq. \eqref{eq:19}, eq.\eqref{eq:18} reduce to 
\begin{equation}
-\frac{\dot{H}}{H^2}=\frac{3}{\alpha+1}=n.\label{eq:20}
\end{equation}
The deceleration parameter $q=-1-\frac{\dot{H}}{H^2}$ for this model becomes $q=\frac{2-\alpha}{\alpha+1}$. It  is worth to mention here that, $q$ plays an important role in the description of the dynamics of the universe. If it is positive it signifies a decelerating universe and for its negative value, the model represents an accelerating universe. Observations confirm that, the universe is in a state of acceleration in the present epoch and hence $q$ should be negative. Observations from type Ia Supernovae predict an accelerating universe with deceleration parameter $q=-0.81\pm 0.14$ in the present time \cite{Rapetti07}. Type Ia Supernovae data in combination with BAO and CMB observations constrain the deceleration parameter as $q=-0.53^{+0.17}_{-0.13}$ \cite{Giostri12}. This behaviour of the universe sets up an allowed range for the parameter $\alpha$ as $\alpha >2$ or $\alpha <-1$. The deceleration parameter $q$ in the present model does not change with time and is decided by the parameter $\alpha$. From eqs.\eqref{eq:15} and \eqref{eq:20}, we can have 
\begin{equation}
R=\chi H^2,\label{eq:21}
\end{equation}
where $\chi=2\xi\left((k+2)n-(k^2+2k+3)\xi\right)$.

Integration of \eqref{eq:19} now yields
\begin{equation}
f(R)=\left[\frac{2}{\chi^{\frac{\alpha}{2}}(\alpha +2)}\right]R^{1+\frac{\alpha}{2}}.\label{eq:22}
\end{equation}
It is obvious that, for $\alpha =0$, the model reduces to that of General Relativity. One should note that, the exponent of $R$ in the function $f(R)$ is not affected by the anisotropic parameter $k$ and is only decided by $\alpha$. However, the functional $f(R)$ is affected by the anisotropy of the model through the coefficient $\frac{2}{\chi^{\frac{\alpha}{2}}(\alpha +2)}$. The anisotropic effect is controlled by the factor $\chi$. 

Integrating eq.\eqref{eq:20}, we obtain the Hubble parameter as
\begin{equation}
H=\frac{H_0}{n(t-t_0)+1},\label{eq:23}
\end{equation} 
and consequently the scale factor as

\begin{equation}
a=a_0\left[n(t-t_0)+1\right]^{\frac{H_0}{n}}.\label{eq:23a}
\end{equation}
Here, $H_0$ and $a_0$ are respectively the Hubble parameter and scale factor at the present epoch $t_0$. Since the universe is expanding with time, we must have $\frac{H_0}{n}>0$, which constrains $\alpha$ to be positive. Now we are in a  position to eliminate the possibility that $\alpha <-1$ and can conclude in favour of the range $\alpha >2$. 

\subsubsection{Cosmic jerk}
The jerk parameter defined as $j=\frac{\dddot{a}}{aH^3}$ is a dimensionless parameter based on the third derivative of the scale factor and therefore provides a good description of the model involving the geometry of the universe.  It can also be expressed in terms of the deceleration parameter $q$ as $j=q+2q^2-\frac{\dot{q}}{H}$. The cosmic dynamics of the present universe requires a positive value of the jerk parameter and a negative value of the deceleration parameter. The jerk parameter for $\Lambda$CDM model is $1$. However, the value of jerk parameter as  obtained from the combination of three kinematical data sets comprising gold sample of type Ia Supernova \cite{Reiss2004}, data from SNLS project \cite{Astier06} and the X-ray galaxy cluster distance measurements \cite{Rapetti07} is $j=2.16^{+0.81}_{-0.75}$. In the present model, the deceleration parameter comes out to be  a constant quantity and therefore, $j=q(1+2q)$. Since $q=\frac{2-\alpha}{\alpha+1}$, the jerk parameter for this model can be obtained as $j=1-9\frac{\alpha-1}{(\alpha+1)^2}$. The value of jerk parameter from the combined data set limits the value of $\alpha$ in the range $0.37\leq \alpha \leq 0.81$. Even if this range of values of $\alpha$ provides a positive value of $j$, it fails to reproduce a negative deceleration parameter and therefore it does not provide us an accelerating model for the present case of $f(R)$. It is worth to mention here that, the determination of the jerk parameter and deceleration parameter involves the measurements of Supernova of redshift $z \geq 1$ which is a formidable task and the derived range of $\alpha$ is not totally conclusive.

\subsubsection{Stability analysis}
The $f(R)$ Lagrangian cannot be arbitrarily chosen but it must satisfy some constraints from observations as well as theoretical justification. A cosmological model in $f(R)$ gravity should be stable and be able to mimic a universe consistent with observations. Also, the cosmological models should satisfy local gravity tests such as the constraint from Solar system. Stable cosmological models in $f(R)$ gravity require that
\begin{equation}
f_R>0,\label{eq:24}
\end{equation}
and
\begin{equation}
f_{RR}>0,\label{eq:25}
\end{equation}
where $f_{RR}=\frac{d^2f(R)}{dR^2}$. In the present model, $H$ is positive and consequently $f_R=H^{\alpha}>0$. Besides the above two constraints, the Solar system constraints restricts the value of $f_R$ in present universe to be $|f_{R0}|< < 1$. In fact. as per the calculation of Hu and Sawicki \cite{Hu07}, the General Relativity results within the Solar system can be recovered for $|f_{R0}|<10^{-6}$.

From \eqref{eq:22}, we obtain $f_{RR}=\frac{\alpha}{2R}$. $\alpha$ is constrained to be positive and consequently, the second condition of stability in eq. \eqref{eq:25} requires that the Ricci scalar $R$ should be positive. It is obvious from eq.\eqref{eq:21} that, the second stability can be satisfied in the present model for $\chi >0$.

\subsection{Model-II}
We may consider another choice for $f_R$ in the form
\begin{equation}
\frac{\dot{f_R}}{f_R}=\beta H, \label{eq:26}
\end{equation}
so that eq. \eqref{eq:18} becomes 
\begin{equation}
\dot{H}+(3+\beta)H^2=0.\label{eq:27}
\end{equation}
Here $\beta$ is an arbitrary constant. The deceleration parameter for this models becomes, $q=2+\beta$. As in the previous model, the deceleration parameter is not evolving with time and its nature is decided by the value of the parameter $\beta$. In order to get viable accelerating models, the parameter $\beta$ must be having a value less than $-2$ i.e $\beta <-2$.

Integrating eq. \eqref{eq:27}, we get
\begin{equation}
H=\frac{H_0}{(3+\beta)(t-t_0)+1},\label{eq:28}
\end{equation} 
and consequently the scale factor as

\begin{equation}
a=a_0\left[(3+\beta)(t-t_0)+1\right]^{\frac{H_0}{3+\beta}}.\label{eq:29}
\end{equation}
It is clear from the expression of the scale factor that, for an expanding model we require $\beta$ to be greater than $-3$ i.e. $\beta >-3$. In otherwords, for an accelerating universe as has been confirmed from  a lot of observational data, the parameter $\beta$ can be constrained in the range $-3<\beta<-2$. 

Integration of eq. \eqref{eq:26} alongwith eq. \eqref{eq:28} provides 
\begin{equation}
f_R=k_1H^{-\frac{\beta H_0}{3+\beta}},\label{eq:30}
\end{equation}
where, $k_1 \simeq H_0^{\frac{\beta H_0}{3+\beta}}$. The Ricci scalar for this model can be expressed as $R=\chi_1 H^2$, where $\chi_1=2\xi\left[(k+2)(3+\beta)-(k^2+2k+3)\xi\right]$. The function $f(R)$ is obtained to be
\begin{equation}
f(R)=k_1\left(\frac{R}{\chi_1}\right)^{-\frac{\beta H_0}{2(3+\beta)}}.\label{eq:31}
\end{equation}
In this model also, the exponent of $R$ in the expression of $f(R)$ is not affected by the anisotropic parameter $k$. However, the anisotropy in the expansion rates affects the function $f(R)$ through the factor $\chi_1$. 

The jerk parameter for this model is derived as $j=2\beta^2+9\beta+10$. The observational constraints on $j$ restrict the parameter $\beta$ in the range $-3.494\leq \beta \leq -1.006$ which is well within the limits already set from the analysis of the deceleration parameter. However, if we consider the mean value of $j$ as $2.16$, the allowed values of $\beta$ can be $-1.181$ and $-3.319$.

We can test the stability of this model. Since $f(R)>0$, the model satisfies the first stability condition. In order to test the second condition, from eq. \eqref{eq:31}, we obtain 
\begin{equation}
f_{RR}=-(k_1)^{\frac{\beta H_0}{2(3+\beta)}}\left[\frac{\beta H_0}{2\chi_1(3+\beta)}\right]\left(f_R\right)^{1-\frac{\beta H_0}{2(3+\beta)}}.\label{eq:32}
\end{equation}
Since the parameter $\beta$ is constrained to be in the range $-3<\beta<-2$, we require a positive value of $\chi_1$ to satisy the second stability condition $f_{RR}>0$. 

\section{Conclusion}
In the present work, we have constructed some anisotropic models in the framework of a modified theory of gravity dubbed as $f(R)$ gravity. The reconstruction of the models are based upon vacuum solutions to the field equations. Plane symmetric LRSBI models are considered to incorporate anisotropy in the expansion of rates along different spatial directions. The field equations in the modified gravity theory are simplified by assuming that the shear scalar is proportional to the scalar expansion. This assumption helps maintain an anisotropic relationship through out the cosmic evolution among the directional Hubble rates and the directional scale factors. Also with this consideration, we are able to describe the dynamics of the LRSBI universe in $f(R)$ gravity in a general form. In General Relativity, such an assumption leads to a decelerating universe. However, it is shown that, the functional $f(R)$ else than $f(R)=R$ modifies the situation and contributes for cosmic acceleration. The additional term in the simplified field equation  can be conceived as the contribution coming from dark energy components or that of the scalar fields.

Two classes of anisotropic cosmological models are constructed by adopting a different method to chose the functional form for $f(R)$. From the constructed models, we obtained the exact forms for the functional $f(R)$. The model parameters are constrained from certain physical basis. The viability of the models are tested through the stability analysis and the conditions of the stability are discussed. In the models, it is shown that, the anisotropic parameter does not affect the power of the Ricci scalar $R$ in the functional $f(R)$. However, the function $f(R)$ is affected by the anisotropic parameter through a multiplicative coefficient term. The analysis shows that, even if, the role of anisotropic is not that much influential in deciding the trend of the models in $f(R)$ gravity, but it has certain importance in providing viable accelerating models in this modified framework.
\section{Acknowledgement}
BM acknowledges SERB-DST, New Delhi, India for financial support to carry out the Research project [No.-SR/S4/MS:815/13]. SKT acknowledges the hospitality of BITS, Pilani, Hyderabad Campus(India) during an academic visit where a part of this work is done.

\end{document}